\documentstyle[15lomcon,cite,epsfig]{article}
\begin{document}

\title{RARE {\boldmath $B$}-MESON DECAYS \footnote{
Talk presented at the 15th Lomonosov Conference on Elementary Particle
Physics, Moscow, Russia, August 18-24th, 2011.}}

\author{ Miko{\l}aj Misiak}

\address{Institute of Theoretical Physics, Faculty of Physics, 
         University of Warsaw,\\ ul. Ho\.za 69, 00-681 Warsaw, Poland}

       \maketitle\abstracts{Rare decays of the $B$-meson that arise due to
         loop-mediated FCNC transitions are known to provide important
         constraints on beyond-SM theories. Basic properties of several such
         decays are reviewed here.}
\section{Introduction}
Flavour Changing Neutral Current (FCNC) phenomena arise at the one-loop level
in the Standard Model (SM). They may receive similar loop contributions from
beyond-SM particles. Many rare decays of $B$ mesons belong to this class of
processes. I will begin with discussing $B_s \to \mu^+ \mu^-$ that receives
particular attention this year. Next, I will pass to other decay modes that
are generated by the quark-level $b \to s \gamma$ and $b \to s l^+ l^-$
transitions.
\section{{\boldmath $B_s \to \mu^+ \mu^-$} --- the 2011 highlight}
The decay of $B_s$ to two muons has a very clean experimental signature -- a
sharp peak in the dimuon invariant mass. However, its branching ratio in the
SM is extremely small (see Sec.~\ref{sec:more}):
\begin{equation} \label{bmmSM}
{\cal B}(B_s \to \mu^+ \mu^-)_{\rm SM} = (3.34 \pm 0.21) \times 10^{-9}.
\end{equation}
It is known to be very sensitive to new physics even in models with Minimal
Flavour Violation~\cite{D'Ambrosio:2002ex}.  Enhancements by orders of
magnitude are possible even when constraints from all the other available
observables are taken into account. In July 2011, 
a new bound on the branching ratio was announced by the CDF
Collaboration~\cite{Aaltonen:2011fi}:
${\cal B}(B_s \to \mu^+ \mu^-)_{\rm CDF} < 40 \times 10^{-9}~~\mbox{at 95\% C.L.}$
Since an excess of signal events remained after cuts, their measurement
could have also been interpreted as an observation:
${\cal B}(B_s \to \mu^+ \mu^-)_{\rm CDF} = \left( 18^{+11}_{-9} \right) \times 10^{-9}.$
An excitement about its large central value ended two weeks later at the
EPS 2011 conference 
where the LHCb and CMS collaborations announced results of their
searches. They observed no signal excess and presented upper bounds only, whose
combination reads~\cite{BPH-11-019-PAS}
\begin{equation} \label{LHCbound}
{\cal B}(B_s \to \mu^+ \mu^-)_{\rm LHC} < 10.8 \times 10^{-9}~~\mbox{at 95\% C.L.}
\end{equation}
At present (November 2011), the LHC experiments have accumulated data samples
that are several times larger than those used for EPS 2011. Updates of their
analyses are eagerly awaited.
\section{The low-energy effective Lagrangian}
Before continuing, let me recall the basic framework that is used for
calculations of flavour-changing observables at scales much below the
electroweak one.  We pass from the full theory of electroweak interactions to
an effective one by removing the high-energy degrees of freedom,
i.e. integrating out the $W$-boson and all the other particles with masses of
order $M_W$ or heavier. The resulting Lagrangian takes the form
\begin{equation} \label{Leff}
{\cal L}_{\rm eff} = {\cal L}_{{\rm QCD}\times{\rm QED}}({\rm leptons},{\rm quarks}\neq t)
+ N \sum_n C_n Q_n,
\end{equation}
where $Q_n$ are higher-dimensional interaction terms (operators), $C_n$ are
the corresponding coupling constants (Wilson coefficients), and $N$ is a
normalization constant. Information on electroweak-scale physics is encoded in
the values of $C_i(\mu)$. Such an effective theory is a modern version of the
Fermi theory for weak interactions. It is ``non-renormalizable'' in the
traditional sense, but actually renormalizable because an infinite set of
operators of arbitrarily high dimensions is included. It is also predictive,
because all the $C_i$ are calculable, and only a finite number of them is
necessary at each given order in the (external momenta)/$M_W$ expansion. The
main advantages of using the effective theory language are easier account for
symmetries and the possibility of resumming large logarithms like
$\left(\alpha_s \ln M_W^2/\mu^2 \right)^n$ from all orders of the perturbation
series using renormalization group techniques.
\section{More on {\boldmath $B_s \to l^+ l^-$} and {\boldmath $B^0 \to l^+
    l^-$} \label{sec:more}}
There are three dimension-six operators in ${\cal L_{\rm eff}}$ that matter
for $B_s \to \mu^+ \mu^-$ in the SM and beyond. They read
\begin{equation}
Q_A = \left( {\bar b} \gamma_\alpha \gamma_5 s \right) \left( {\bar\mu} \gamma^\alpha \gamma_5 \mu \right),\hspace{5mm}
Q_S = \left( {\bar b} \gamma_5 s \right) \left( {\bar\mu} \mu \right),\hspace{5mm}
Q_P = \left( {\bar b} \gamma_5 s \right) \left( {\bar\mu} \gamma_5 \mu \right).
\end{equation}
Setting
$N = V_{tb}^* V_{ts}\, G_F^2 M_W^2/\pi^2$ 
in Eq.~(\ref{Leff}), one obtains 
\begin{equation} \label{bmmgen}
{\cal B}({\bar B}_s \to \mu^+ \mu^-) = 
\frac{ |N|^2 M^3_{B_s} f^2_{B_s}}{8 \pi\, \Gamma_{\!B_s}} \sqrt{1\!-\!r^2} 
\left[ \left|r C_A - u C_P\right|^2 + \left|u C_S\right|^2 \left(1\!-\!r^2\right) \right]\!,\!
\end{equation}
where $r = 2m_\mu/M_{B_s}$ and $u = M_{B_s}/(m_b+m_s)$. The 
decay constant $f_{B_s}$ parametrizes the matrix element
$\langle 0 | {\bar b} \gamma^\nu \gamma_5 s | B_s(p) \rangle = i p^\nu f_{B_s}$.
Only the coefficient $C_A$ matters in the SM because $C_{S,P} \sim m_\mu/M_W$,
and their effects on the r.h.s. of Eq.~(\ref{bmmgen}) are thus suppressed by
$M_{B_s}^2/M_W^2$ with respect to those of $C_A$. At the leading order,
$C_A^{\rm SM, LO} = \frac{3x^2}{16(x-1)^2}\ln x + \frac{x^2-4x}{16(x-1)}$~
with $x = m_t^2/M_W^2$. When the $\overline{\rm MS}$ mass
$\overline{m}_t(\overline{m}_t) \simeq 165(1)\,$GeV is used in $x$, the ${\cal
  O}(\alpha_s)$ corrections~\cite{Buchalla:1993bv} enhance the branching ratio
by around $+2.2\%$, while the electroweak corrections to $C_A^{\rm SM}$ that
have been calculated in Refs.~\cite{Buchalla:1997kz,Bobeth:2003at} act in the
opposite way, and suppress the branching ratio by around $-1.7\%$.  The
central value in Eq.~(\ref{bmmSM}) has been obtained for $|V_{cb}| =
0.04185(73)$~\cite{Asner:2010qj}, $\tau_{B_s} =
1.472(26)\,$ps~\cite{Nakamura:2010zzi}, and $f_{B_s} =
225(4)\,$MeV~\cite{McNeile:2011ng}. If $f_{B_s} =
242.0(9.5)\,$~MeV~\cite{Bazavov:2011aa} was used instead, the SM result in Eq.~(\ref{bmmSM})
would become $(3.86 \pm 0.36 ) \times 10^{-9}$.

Useful phenomenological expressions for all the $B_s \to l^+ l^-$ and $B^0 \to
l^+ l^-$ branching ratios in the SM can be found in Eqs.~(127)--(132) of
Ref.~\cite{Buchalla:2008jp}.\footnote{
Note different normalization conventions for the operators and their
Wilson coefficients there.}
The quoted uncertainties there should be understood to include around 3\% ones
due to the unknown ${\cal O}(\alpha_s^2)$ and subleading electroweak
corrections.  For the $B_s \to l^+ l^-$ decays, the current experimental
bounds are above the SM predictions by factors ${\cal O}(10^6)$, $3.3$, ${\cal
  O}(10^5)$, for $l=e,\;\mu,\;\tau$, respectively.  The corresponding numbers
for $B^0 \to l^+ l^-$ are ${\cal O}(10^7)$, 35, ${\cal O}(10^5)$. Thus, the
muonic decay of $B_s$ is definitely the most restrictive at present.

Constraints on the Two-Higgs-Doublet Model II from Fig.~3 of
Ref.~\cite{Logan:2000iv} can easily be updated to include the new upper bound
on $B_s \to \mu^+ \mu^-$ (\ref{LHCbound}) and the lower bound
$M_{H^\pm} > 295\;$GeV that comes from $\bar B \to X_s \gamma$
\cite{Misiak:2006zs}. It follows that $\tan\beta < 50$ remains allowed for the
charged Higgs boson mass values that survive the $\bar B \to X_s \gamma$
constraint.

As far as the Minimal Supersymmetric Standard Model (MSSM) is concerned, the
first analysis~\cite{Akeroyd:2011kd} performed after the EPS 2011 conference
implies that $\tan\beta$ larger than 50 is hard to accommodate in the CMSSM
given the current $B_s \to \mu^+ \mu^-$ bounds. Assuming SM-like measurement
with $\pm 10\%$ accuracy, the authors find that $\tan\beta$ must be smaller
than around 40 for stop $\tilde{t}_1$ masses up to 2$\,$TeV.
\section{What other rare {\boldmath $B$} decays are interesting?}
There are two basic scenarios for flavour physics beyond the SM.  The scenario
"{\bf A}" ({\bf A}ttractive or {\bf A}rbitrary) is characterized by Generic
Flavour Violation in interactions of new particles with the SM ones. Its
properties are as follows: 
{\it (i)} Large deviations from the SM in the Wilson coefficients are
possible.
{\it (ii)} Observable new physics effects may arise despite QCD-induced theory
uncertainties in many FCNC decays of the $B$ meson, like penguin-induced
exclusive hadronic decays, $B \to K^* \gamma$, $B \to K^{(*)} l^+ l^-$, etc.
{\it (iii)} Interesting constraints can be obtained from branching ratios,
angular distributions and various asymmetries.

The scenario "{\bf B}" ({\bf B}oring or {\bf B}eautiful) corresponds to quite
heavy new particles and Minimal Flavour Violation. In such a case: 
{\it (i)}~Only mild beyond-SM effects in most of the Wilson coefficients are expected.
{\it (ii)}~CP-asymmetries are unaffected.  
{\it (iii)}~Precise measurements are needed. Consequently, small rates are 
            not welcome, i.e. $b \to s$ transitions are preferred over $b \to d$ ones. 
{\it (iv)}~Precise theory predictions in the SM case are needed, which implies
           that inclusive rather than exclusive hadronic final states are preferred.
{\it (v)}~Suppression in the SM due to parameters other than CKM
          angles is a positive property of any considered observable because it
          increases sensitivity to new physics.
          {\it (vi)}~Apart from $B \to l^+ l^-$, the inclusive decay $\bar B
          \to X_s \gamma$ is of main interest.  Other inclusive decays like
          $\bar B \to X_s \nu \bar\nu$ or $\bar B \to X_s l^+ l^-$ undergo
          no chiral suppression in the SM but still deserve consideration.
{\it (vii)}~Exclusive observables (like asymmetries) may still be useful to
            resolve discrete ambiguities.

In the following, I shall comment on several observables that remain relevant in the case ``{\bf B}''.
\section{\boldmath $\bar B \to X_s \gamma$}
The inclusive decay rate $\Gamma(\bar B \to X_s \gamma)$ with $\bar B = \bar
B^0$ or $B^-$ and the lower cut on the photon energy $E > E_0$ is well
approximated by the corresponding perturbative partonic rate $\Gamma(b \to
X_s^p \gamma)$, provided $E_0$ is neither too large, nor too small. For a
conventional choice $E_0 = 1.6\,{\rm GeV} \simeq m_b/3$, unknown
non-perturbative corrections to this approximation have been analyzed in
detail in Ref.~\cite{Benzke:2010js}, and estimated to remain at around $\pm
5\%$ level. The goal of the ongoing perturbative calculations (see
Ref.~\cite{Misiak:2010dz} for a review) is to make the ${\cal O}(\alpha_s^2)$
uncertainties negligible with respect to the non-perturbative ones. At
present, the SM prediction
${\cal B}(\bar B \to X_s \gamma)^{\rm SM} = \left( 3.15 \pm 0.23 \right)
\times 10^{-4}$~\cite{Misiak:2006zs}
agrees with the world average
${\cal B}(\bar B \to X_s \gamma)^{\rm exp} = \left( 3.55 \pm 0.24 \pm 0.09\right)
\times 10^{-4}$~\cite{Asner:2010qj}
within $1.2\sigma$. This fact has been used to derive constraints on various
new physics models, like the bound on $M_{H^\pm}$ that has been mentioned in
Sec.~\ref{sec:more}, or effects in the recent MSSM parameter space
fits~\cite{Buchmueller:2011sw}.

\section{Processes generated by the quark-level {\boldmath $b \to s l^+ l^-$} decay}

Contrary to $\bar B \to X_s \gamma$ and $B_{(s)} \to l^+ l^-$, the quark-level
$b \to s l^+ l^-$ decay undergoes no chiral suppression in the SM, which makes
it less sensitive to new physics. It is also more complicated due to partial
screening of beyond-SM effects by $J/\psi$ and higher $c\bar c$ resonances in
the dilepton spectrum. A very recent model-independent analysis of observables
that are available in processes generated by this decay has been presented in
Ref.~\cite{Altmannshofer:2011gn}.  The authors consider inclusive $\bar B \to
X_s l^+ l^-$ in various regions of the dilepton invariant mass, asymmetries of
angular distributions in $B \to K^* l^+ l^-$, as well as the branching ratio
and CP asymmetry in the radiative mode. No significant (larger than $2\sigma$)
deviations from the SM are found. However, allowed regions in the Wilson
coefficient space remain large, so there is no clear indication which scenario
(``{\bf A}'' or ``{\bf B}'') is preferred.
\section{Summary}
Rare $B$ decays provide improving constraints on beyond-SM physics, with a
prominent role played by $B_s \to \mu^+ \mu^-$ this year. New results are
awaited soon.
\section*{Acknowledgments}
This work has been supported in part by the Ministry of Science and Higher
Education (Poland) as research project N~N202~006334 (2008-11), and by the DFG
through the ``Mercator'' guest professorship program.

\end{document}